\documentclass[12pt]{article}
\PassOptionsToPackage{hyphens}{url}
\usepackage[pagebackref=true,colorlinks]{hyperref}
\usepackage{amsmath,amssymb,amsthm,mathtools,amsrefs,mathrsfs,marvosym}
\usepackage[utf8]{inputenc}
\usepackage[T1]{fontenc}
\usepackage{pifont}
\usepackage{mathpazo}
\usepackage{authblk} 
\usepackage{eulervm}
\usepackage{csquotes}
\usepackage{bbm}
\usepackage{lscape}
\usepackage{etex} 
\usepackage[table]{xcolor}
\usepackage{tikz}
\usetikzlibrary{arrows.meta}
\usetikzlibrary{decorations.pathmorphing,shapes}
\usetikzlibrary{calc}
\usepackage{comment}
\usepackage{adjustbox}
\usepackage{scalerel,stackengine}
\usepackage{stmaryrd}
\usepackage{fancyhdr}
\usepackage{soul}
\usepackage{dsfont}
\usepackage[normalem]{ulem}
\usepackage{longtable}
\usepackage[bottom]{footmisc}
\usepackage{graphicx}
\usepackage{caption}
\usepackage{subcaption}
\usepackage{pictexwd, dcpic}
\usepackage{float}
\usepackage{enumerate}
\usepackage[all]{xy} 
\usetikzlibrary{circuits.ee.IEC}
\hypersetup{
	colorlinks=true,
    linktoc=all,
    linkcolor=blue,  
    citecolor=blue,
    }

\usepackage{tocbasic}
\DeclareTOCStyleEntry[
  beforeskip=-0.3em plus 1pt,
  pagenumberformat=\textbf
]{tocline}{section}

\makeatletter
\newcommand*{\shifttext}[2]{%
  \settowidth{\@tempdima}{#2}%
  \makebox[\@tempdima]{\hspace*{#1}#2}%
}
\makeatother
\makeatletter
\renewcommand*\env@matrix[1][\arraystretch]{%
  \edef\arraystretch{#1}%
  \hskip -\arraycolsep
  \let\@ifnextchar\new@ifnextchar
  \array{*\c@MaxMatrixCols c}}
\makeatother

\stackMath
\newcommand\reallywidehat[1]{%
\savestack{\tmpbox}{\stretchto{%
  \scaleto{%
    \scalerel*[\widthof{\ensuremath{#1}}]{\kern.1pt\mathchar"0362\kern.1pt}%
    {\rule{0ex}{\textheight}}
  }{\textheight}%
}{2.4ex}}%
\stackon[-6.9pt]{#1}{\tmpbox}%
}
\makeatletter
\pgfmathdeclarefunction{erf}{1}{%
  \begingroup
    \pgfmathparse{#1 > 0 ? 1 : -1}%
    \edef\sign{\pgfmathresult}%
    \pgfmathparse{abs(#1)}%
    \edef\x{\pgfmathresult}%
    \pgfmathparse{1/(1+0.3275911*\x)}%
    \edef\t{\pgfmathresult}%
    \pgfmathparse{%
      1 - (((((1.061405429*\t -1.453152027)*\t) + 1.421413741)*\t 
      -0.284496736)*\t + 0.254829592)*\t*exp(-(\x*\x))}%
    \edef\y{\pgfmathresult}%
    \pgfmathparse{(\sign)*\y}%
    \pgfmath@smuggleone\pgfmathresult%
  \endgroup
}
\makeatother

\theoremstyle{theorem}
\newtheorem{theorem}[equation]{Theorem}
\newtheorem{lemma}[equation]{Lemma}
\newtheorem{proposition}[equation]{Proposition}
\newtheorem{corollary}[equation]{Corollary}
\theoremstyle{definition}
\newtheorem{definition}[equation]{Definition}
\newtheorem{construction}[equation]{Construction}
\newtheorem{question}[equation]{Question}

\newtheorem{problem}[equation]{Problem}
\newtheorem{example}[equation]{Example}
\newtheorem{exercise}[equation]{Exercise}
\newtheorem*{answer}{Answer}
\newtheorem*{solution}{Solution}
\newtheorem{remark}[equation]{Remark}

\newtheorem{notation}[equation]{Notation}
\newtheorem{noterm}[equation]{Notation and Terminology}

\newcommand\define[1]{\emph{\textbf{#1}}}

\numberwithin{equation}{section}

\newcommand{\be}{\begin{equation}}
\newcommand{\ee}{\end{equation}}
\def\ba{\begin{align}} 
\def\ea{\end{align}}
\newcommand{\bea}{\begin{eqnarray}}
\newcommand{\eea}{\end{eqnarray}}
\newcommand{\bx}{\begin{example}}
\newcommand{\ex}{\end{example}}
\newcommand{\bex}{\begin{exercise}}
\newcommand{\eex}{\end{exercise}}
\newcommand{\ban}{\begin{answer}}
\newcommand{\ean}{\end{answer}}
\newcommand{\bt}{\begin{theorem}}
\newcommand{\et}{\end{theorem}}
\newcommand{\bc}{\begin{corollary}}
\newcommand{\ec}{\end{corollary}}
\newcommand{\blem}{\begin{lemma}}
\newcommand{\elem}{\end{lemma}}
\newcommand{\bp}{\begin{problem}}
\newcommand{\ep}{\end{problem}}
\newcommand{\bn}{\begin{proposition}}
\newcommand{\en}{\end{proposition}}
\newcommand{\bd}{\begin{definition}}
\newcommand{\ed}{\end{definition}}
\newcommand{\bcon}{\begin{construction}}
\newcommand{\econ}{\end{construction}}
\newcommand{\bq}{\begin{question}}
\newcommand{\eq}{\end{question}}
\newcommand{\bprf}{\begin{proof}}
\newcommand{\eprf}{\end{proof}}
\newcommand{\br}{\begin{remark}}
\newcommand{\er}{\end{remark}}
\newcommand{\bs}{\begin{solution}}
\newcommand{\es}{\end{solution}}
\newcommand{\beqs}{\begin{eqnarray}}
\newcommand{\eeqs}{\end{eqnarray}}
\newcommand{\bnt}{\begin{noterm}}
\newcommand{\ent}{\end{noterm}}
\newcommand{\bnot}{\begin{notation}}
\newcommand{\enot}{\end{notation}}

 \let\ov=\overline

\newcommand{\id}{\mathrm{id}}

\newcommand{\lra}{\longrightarrow}

\newcommand{\tr}{{\rm tr} }

\def\F{{{\mathcal{F}}}}

\def\C{{{\mathbb C}}}

\def\N{{{\mathbb N}}}

\newcommand{\Ad}{\mathrm{Ad}}

\def\invexcl{\rotatebox[origin=c]{180}{$!$}}
\newcommand{\bloom}{\operatorname{\invexcl}}

\def\A{\Alg{A}}
\def\B{\Alg{B}}
\def\C{\Alg{C}}
\def\E{\mathcal{E}}
\def\F{\mathcal{F}}

\newcommand{\stoch}{\;\xy0;/r.25pc/:(-3,0)*{}="1";(3,0)*{}="2";{\ar@{~>}"1";"2"|(1.06){\hole}};\endxy\!}
\newcounter{sarrow}

\newcounter{sqarrow}

\DeclareFontFamily{OT1}{pzc}{}
\DeclareFontShape{OT1}{pzc}{m}{it}{ <-> s*[1.2] pzcmi7t }{}
\DeclareMathAlphabet{\mathpzc}{OT1}{pzc}{m}{it}
\newcommand{\Alg}[1]{\mathpzc{#1}}

\newcommand{\ben}{\renewcommand{\theenumi}{\alph{enumi}} 
\renewcommand{\labelenumi}{(\theenumi)}\begin{enumerate}}
\newcommand{\een}{\end{enumerate}}
\title{General covariance for quantum states over time}
\author[$\spadesuit$]{James Fullwood}
\affil[$\spadesuit$]{School of Mathematical Sciences, Shanghai Jiao Tong University, 800 Dongchuan Road, Shanghai, China}
\date{}                     
\begin{document}
\emergencystretch 2em

\maketitle

\begin{abstract}  
The theory of quantum states over time provides an approach to the dynamics of quantum information which is in direct analogy with spacetime and its relation to classical dynamics. In this work, we further such an analogy by formulating a notion of general covariance for the theory of quantum states over time. We then associate a canonical state over time with a density operator which is to evolve under a sequence of quantum processes modeled by completely positive trace-preserving (CPTP) maps, and we show that such a canonical state over time satisfies such a notion of covariance. We also show that the dynamical quantum Bayes' rule transforms covariantly with respect to states over time, and we conclude with a discussion of what it means for a physical law to be generally covariant when formulated in terms of quantum states over time.
\end{abstract}

\vspace{-7mm}
\tableofcontents

\section{Introduction}

There is a prevailing viewpoint that fundamental physics should be based on a primitive notion of information. The theory of quantum states over time is a nascent approach to quantum theory that stems directly from such a viewpoint, formulating the dynamics of quantum information in a way which is directly analogous to spacetime and its relation to classical dynamics \cite{HHPBS17,FuPa22,FuPa22a,Fu23,LiNg23}. As opposed to a density operator which only encodes spatial correlations in quantum systems, quantum states over time encode spatio-\emph{temporal} correlations which arise as a result of a quantum system evolving under quantum processes modeled by completely positive trace-preserving (CPTP) maps. And similar to how spacetime is a single object encompassing the dynamical evolution of space, a quantum state over time is a single operator encoding the dynamical evolution of a quantum state. As such, quantum states over time serve as a quantum-informational analog of spacetime, and in analogy with the Lorentzian signature of the spacetime metric, spatial and temporal correlations combine to yield quantum states over time which are not positive in general (though still self-adjoint). Furthermore, in analogy with how Cauchy hypersurfaces of spacetime correspond to space at a fixed point in time, the reduced operators of a quantum state over time yield genuine density operators, representing the state of the associated quantum system at a fixed point in time. 

We emphasize here that "time" in the theory of quantum states over time is fundamentally discrete, and is viewed as an emergent property associated with the correlations which result as a consequence of a quantum system evolving according to general CPTP dynamics. As such, a continuous universal time parameter as that which appears in the Schr\"odinger equation is then viewed as a smooth approximation to an underlying discrete flow of time which is intrinsic to the dynamical evolution of the system of interest.  

While the study of quantum states over time is still in its formative stages, viewing quantum dynamics through the lens of states over time has already led to insights into the nature of time reversal in quantum theory \cite{FuPa22a}, quantum Bayesian inference and retrodiction \cite{LeSp13,FuPa22a,PaRuBayes}, virtual quantum broadcasting \cite{FPBC23}, and dynamical measures of quantum information \cite{FuPa23}. Quantum states over time have also been shown to coincide with the pseudo-density matrices associated timelike separated Pauli measurements performed on systems of qubits \cite{FJV15,LQDV23}. 

In this work, we further extend the analogy between spacetime and quantum states over time by formulating an analog of general covariance for the theory of quantum states over time. As general covariance in the context of general relativity is the requirement that the dynamical equations of the theory should be diffeomorphism invariant, we take general covariance in the context of quantum theory to be the requirement that dynamical equations governing quantum systems should be \emph{unitarily} invariant. To formulate general covariance for states over time, we then associate a canonical quantum state over time with an initial density operator which is to evolve according to the Liouville-von Neumann equation, and analyze how such a state over time transforms in accordance with a unitary transformation of the algebra of operators containing the initial state and the associated Hamiltonian. Such analysis then yields a natural notion of covariance for quantum states over time which generalizes the covariance property for virtual broadcasting introduced in Ref.~\cite{FPBC23}. Moreover, we show that such a notion of covariance for quantum states over time holds for systems evolving according to general CPTP dynamics, as opposed to strictly unitary transformations as determined by the Liouville-von Neumann equation. 

The present paper is organized as follows. In Section~\ref{SX2}, we lay the mathematical foundation for the study of states over time, providing all necessary definitions and setting the notational conventions. In Section~\ref{QSXT787} we formally introduce quantum states over time associated with general CPTP dynamics, and show that there is a canonical "propagating" system of quantum states over time (Theorem~\ref{PGPGTXS757}). In Section~\ref{SX4}, we formulate a precise notion of general covariance for quantum states over time, and show that the canonical propagating system of quantum states over time is indeed covariant (Theorem~\ref{MTX87}). In Section~\ref{SX5}, we show that the dynamical Quantum Bayes' rule of \cite{FuPa22a} is also covariant in the sense established in Section~\ref{SX4} (Theorem~\ref{CVQBXR57}). We then make some concluding remarks in Section~\ref{SX6}, before ending with an appendix of some technical results required for the proofs our main theorems.

\section{Preliminaries} \label{SX2}

In this work, all mathematics will take place in the category of finite-dimensional $C^*$-algebras. Any such algebra is $*$-isomorphic to an algebra of block diagonal matrices, which we view as an algebra of linear operators on an abstract Hilbert space. In what follows, the caligraphic letters $\A$, $\B$ \ldots will be used to denote finite-dimensional $C^*$-algebras. 

\bd
Given finite-dimensional $C^*$-algebras $\A$ and $\B$, the set of trace-preserving linear maps from $\A$ to $\B$ will be denoted by $\bold{TP}(\A,\B)$, the set of hermitian-preserving, trace-preserving linear maps from $\A$ to $\B$ will be denoted by $\bold{HPTP}(\A,\B)$, and the set of completely positive, trace-preserving linear maps from $\A$ to $\B$ will be denoted by $\bold{CPTP}(\A,\B)$. The \define{Hilbert-Schmidt adjoint} of a linear map $\E:\A\to \B$ is the map $\E^*:\B\to \A$ uniquely determined by the condition
\[
\tr\left(\E(A)^{\dag}B\right)=\tr\left(A^{\dag}\E^*(B)\right)
\] 
for all $A\in \A$ and $B\in \B$, where $\dag$ is used to denote the $*$-operation in a finite-dimensional $C^*$-algebra.
\ed

\bnot
Given $U\in \A$, the mapping $A\mapsto UAU^{\dag}$ will be denoted by $\text{Ad}_U:\A\to \A$.
\enot

\bd
A self-adjoint element $A\in \A$ of unit trace is said to be a \define{virtual state} in $\A$, and the set of all virtual states in $\A$ will be denoted by $\mathcal{V}(\A)$. A virtual state $A\in \A$ is said to be a \define{state} if and only if $A$ is positive, and the set of all states in $\A$ will be denoted by $\mathcal{S}(\A)$.
\ed

\bd
Given a finite set $X$, the commutative $C^*$-algebra consisting of complex-valued functions on $X$ will be denoted by $\mathbb{C}^X$, which will be referred to as a \define{classical algebra}.
\ed

\br[\bf{Classical states are probability distributions}]
Given a classical algebra $\mathbb{C}^X$, we note that there is a canonical bijective correspondence between states on $\mathbb{C}^X$ and probability distributions on $X$. For every $x\in X$, the state on $\mathbb{C}^X$ which takes the value 1 on $x$ and 0 otherwise will be denoted by $\delta_x$. \hspace{12.97cm} $\triangle$
\er

\br[\bf{Classical channels as CPTP maps}] \label{RM7X1787}
If $X$ and $Y$ are finite sets and $\E:\mathbb{C}^X\to \mathbb{C}^Y$ is a CPTP map, then for all $x\in X$ it follows that there exists elements $\E_{yx}\in [0,1]$ such that
\[
\E(\delta_x)=\sum_{y\in Y}\E_{yx}\delta_y.
\]
Moreover, by the trace-preserving condition on $\E$ it follows that $\sum_{y\in Y}\E_{yx}=1$. As such, the elements $\E_{yx}$ are viewed as conditional probabilities of $y$ given $x$, thus any $\E\in \bold{CPTP}(\mathbb{C}^X,\mathbb{C}^Y)$ is equivalent to a classical channel from $X$ to $Y$. \hspace{8.17cm} $\triangle$
\er

We now define fundamental mappings that will play a prominent role in this work.

\bd
Given a finite-dimensional $C^*$-algebra $\A$, let $\mu_{\A}:\A\otimes \A\to \A$ and $\widetilde{\mu}_{\A}:\A\otimes \A\to \A$ denote the maps corresponding to the linear extensions of the assignments
\[
\mu_{\A}(A_1\otimes A_2)=A_1A_2 \quad \& \quad \widetilde{\mu}_{\A}(A_1\otimes A_2)=A_2A_1.
\]
The \define{canonical broadcasting map} is then the linear map $\mathfrak{B}_{\A}:\A\to \A\otimes \A$ given by
\be \label{CVXB67}
\mathfrak{B}_{\A}=\frac{1}{2}\left(\mu_{\A}^*+\widetilde{\mu}_{\A}^*\right),
\ee
where $\mu_{\A}^*$ and $\widetilde{\mu}_{\A}^*$ denote the Hilbert-Schmidt adjoints of $\mu_{\A}$ and $\widetilde{\mu}_{\A}$.
\ed

\br[\bf{Explicit formula for canonical broadcasting}] 
If $\A$ is a matrix algebra, it was shown in Ref.~\cite{FuPa23} (Lemma~A.3) that for all $A\in \A$,
\[
\mu_{\A}^*(A)=(A\otimes \mathds{1})\tt{SWAP}, 
\]
where ${\tt SWAP}\in \A\otimes \A$ is the unique matrix satisfying ${\tt SWAP}(A_1\otimes A_2){\tt SWAP}=A_2\otimes A_1$ for all $A_1,A_2\in \A$. If we denote the matrix units in $\A$ by $|i\rangle \langle j|$, then 
\[
{\tt SWAP}=\sum_{i,j}|i\rangle \langle j|\otimes |j\rangle \langle i|.
\]
And since $\widetilde{\mu}_{\A}=\mu_{\A}\circ \gamma$ (where $\gamma:\A\otimes \A\to \A\otimes \A$ is the lexicographic swap isomorphism), it follows that $\widetilde{\mu}_{\A}^*=\gamma\circ \mu^*_{\A}$, thus for all $A\in \A$,
\[
\widetilde{\mu}_{\A}^*(A)=\gamma\Big((A\otimes \mathds{1}){\tt SWAP}\Big)={\tt SWAP}\Big((A\otimes \mathds{1}){\tt SWAP}\Big){\tt SWAP}={\tt SWAP}(A\otimes \mathds{1})
\]
where the final equality follows from the fact that $\gamma(A_1\otimes A_2)={\tt SWAP}(A_1\otimes A_2){\tt SWAP}$ for all $A_1,A_2\in \A$. It then follows that the canonical broadcasting map $\mathfrak{B}_{\A}:\A\to \A\otimes \A$ is given by
\be \label{CNCLXBXS671}
\mathfrak{B}_{\A}(A)=\frac{1}{2}\Big\{(A\otimes \mathds{1}),{\tt SWAP}\Big\} \qquad \forall A\in \A,
\ee
where $\{*,*\}$ denotes the anti-commutator. \hspace{9.37cm}$\triangle$
\er

\br[\bf{The Virtual Broadcasting Theorem}]
In Ref.~\cite{FPBC23}, it was shown that in the category of matrix algebras, the canonical broadcasting map $\mathfrak{B}_{\A}:\A\to \A\otimes \A$ is uniquely characterized by the following three conditions:
\begin{itemize}
\item
\underline{Covariance}: For every unitary $U\in \A$, and for all $A\in \A$,
\[
\mathfrak{B}_{\A}(UAU^{\dag})=(U\otimes U)\mathfrak{B}_{\A}(A)(U\otimes U)^{\dag}.
\]
\item
\underline{Permutation Invariance}: For every $A\in \A$,
\[
\mathfrak{B}_{\A}(A)={\tt SWAP}\hspace{0.1mm}\mathfrak{B}_{\A}(A){\tt SWAP}.
\]
\item
\underline{Classical Consistency}: If $\mathcal{D}:\A\to \A$ is the decoherence map given by $\mathcal{D}(|i\rangle \langle j|)=\delta_{ij}|i\rangle \langle i|$, then 
\[
(\mathcal{D}\otimes \mathcal{D})\circ \mathfrak{B}_{\A}\circ \mathcal{D}=\mathfrak{B}_{\text{cl}},
\]
where $\mathfrak{B}_{\text{cl}}$ is the classical broadcasting map on diagonal matrices given by $\mathfrak{B}_{\text{cl}}(|i\rangle \langle i|)=|i\rangle \langle i|\otimes |i\rangle \langle i|$. \hspace{13.83cm} $\triangle$
\end{itemize}
\er

\bd \label{BLXDXF677}
Given a pair $(\A,\B)$ of finite-dimensional $C^*$-algebras, the \define{bloom} of a map $\E\in \bold{TP}(\A,\B)$ is the map $\bloom(\E)\in \bold{TP}(\A,\A\otimes \B)$ given by
\be \label{BLXDXF6777}
\bloom(\E)=(\id_{\A}\otimes \E)\circ \mathfrak{B}_{\A},
\ee
where we recall $\mathfrak{B}_{\A}$ is the canonical broadcasting map given by \eqref{CVXB67}.
\ed

\br[\bf{The bloom map is HPTP}] \label{BXLHPTPX37}
While the map $\bloom(\E)$ is trace-preserving for all $\E\in \bold{TP}(\A,\B)$, it is rarely positive. For example, if $\E=\id_{\A}$ with $\A$ a matrix algebra and $\rho\in \mathcal{S}(\A)$ is a state, then it was shown in Ref.~\cite{FuPa23} that $\bloom(\E)(\rho)$ has negative eigenvalues. However, in Ref.~\cite{Fu23} it was shown that $\bloom(\E)$ is hermitian-preserving if and only if $\E$ is hermitian-preserving, thus $\bloom(\E)\in \bold{HPTP}(\A,\A\otimes \B)$ for all $\E\in \bold{HPTP}(\A,\B)$. \hspace{9.27cm} $\triangle$
\er

\bnot
Given a positive natural number $n\in \N$, we let $[n]=\{1,...,n\}$.
\enot

\bd
Let $(\A_0,...,\A_n)$ be an $(n+1)$-tuple of finite-dimensional $C^*$-algebras. An \define{$n$-chain} consists of an $n$-tuple $(\E_1,...,\E_n)$ such that $\E_{i}:\A_{i-1}\to \A_i$ is a linear map for all $i\in [n]$. The set of all such $n$-chains with $\E_i\in \bold{TP}(\A_{i-1},\A_i)$ for all $i\in [n]$ will be denoted by $\bold{TP}(\A_0,...,\A_n)$, the set of all such $n$-chains with $\E_i\in \bold{HPTP}(\A_{i-1},\A_i)$ for all $i\in [n]$ will be denoted by $\bold{HPTP}(\A_0,...,\A_n)$, and the set of all such $n$-chains with $\E_i\in \bold{CPTP}(\A_{i-1},\A_i)$ for all $i\in [n]$ will be denoted by $\bold{CPTP}(\A_0,...,\A_n)$.
\ed

\bd \label{NBLXMS71}
The \define{bloom} of an $n$-chain $(\E_1,...,\E_n)\in \bold{TP}(\A_0,...,\A_n)$ is defined recursively as the map $\bloom(\E_1,...,\E_n)\in \bold{TP}(\A_0,\A_0\otimes \cdots \otimes \A_n)$ given by
\be \label{BLXDFSXS71}
\bloom(\E_1,...,\E_n)=\bloom(\bloom(\E_2,...,\E_n)\circ \E_1).
\ee
\ed

\br[\bf{A closed form expression for the bloom}] \label{BLXHPTP987}
In Ref.~\cite{Fu23} it was shown that
\be \label{CLXFXM87}
\bloom(\E_1,...,\E_n)=\bloom(\E_n\circ \tr)\circ \cdots \circ \bloom(\E_2\circ \tr)\circ \bloom(\E_1),
\ee
where $\tr$ denotes the universal partial trace map that maps an iterated tensor product onto its right-most factor. In fact, there are actually the $n$th Catalan number $c_n=\frac{1}{n+1}\binom{2n}{n}$ of equivalent expressions for $\bloom(\E_1,...,\E_n)$, corresponding to the $c_n$ ways of completely parenthesizing the iterated tensor product $\A_0\otimes \cdots \otimes \A_n$. And since $\bloom(\E)\in \bold{HPTP}(\A,\A\otimes \B)$ for all $\E\in \bold{HPTP}(\A,\B)$ (see Remark~\ref{BXLHPTPX37}), it follows from \eqref{CLXFXM87} that $\bloom(\E_1,...,\E_n)\in \bold{HPTP}(\A_0,\A_0\otimes \cdots \otimes \A_n)$ for all $(\E_1,...,\E_n)\in \bold{HPTP}(\A_0,...,\A_n)$. \hspace{11.27cm} $\triangle$ 
\er

\section{Quantum states over time} \label{QSXT787}

Given a virtual state $\rho\in \mathcal{V}(\A_0)$ which is to evolve according to an $n$-chain $(\E_1,...,\E_n)\in \bold{CPTP}(\A_0,...,\A_n)$, an associated virtual state $(\E_1,...,\E_n)\star \rho\in \mathcal{V}(\A_0\otimes \cdots \otimes \A_n)$ is said to be a \emph{quantum state over time} if and only if for all $i\in \{0,...,n\}$ we have
\be \label{MGXCDX781}
\tr_{\A_0\otimes \cdots \otimes \hat{\A_i}\otimes \cdots \otimes \A_n}\Big((\E_1,...,\E_n)\star \rho\Big)=\rho_i,
\ee
where $\rho_0=\rho$, $\rho_{i}=(\E_i\circ \E_{i-1}\circ \cdots \circ \E_1)(\rho)$ and $\tr_{\A_0\otimes \cdots \otimes \hat{\A_i}\otimes \cdots \otimes \A_n}:\A_0\otimes \cdots \otimes \A_n\to \A_i$ is the partial trace for all $i\in \{0,...,n\}$. An assignment of the form 
\[
(\E_1,...,\E_n,\rho)\longmapsto (\E_1,...,\E_n)\star \rho
\]
where $(\E_1,...,\E_n)\star \rho$ is a quantum state over time for every $n$-chain $(\E_1,...,\E_n)$ with $n>0$ and for every virtual state $\rho$ is then said to be a \emph{system of quantum states over time}. While we will mostly consider quantum states over time $(\E_1,...,\E_n)\star \rho$ with $\rho$ a state, it will be useful to also consider states over time associated with the dynamical evolution of virtual states as well. We note that while $(\E_1,...\E_n)\star \rho$ is defined to be a virtual state for all $n$-chains $(\E_1,...,\E_n)$ and virtual states $\rho$, there is nothing in the definition of quantum state over time that ensures that $(\E_1,...,\E_n)\star \rho$ is in fact positive, even if $\rho$ is a state (as opposed to a virtual state).

As a quantum state over time $(\E_1,...,\E_n)\star \rho$ contains the data of the individual states $\rho_i$ which result throughout the dynamical process associated with the $n$-chain $(\E_1,...,\E_n)$, such a state over time is meant to serve as a quantum analog of a fibration $\varphi:S\to [t_0,t_1]$ of a local patch $S$ of spacetime into spatial fibers $S_t=\varphi^{-1}(t)$ for all $t\in [t_0,t_1]$. However, there is no continuous time parameter associated with quantum states over time, as time occurs in discrete time-steps according to the $n$-chain $(\E_1,...,\E_n)$, which one may view as a quantization of the interval $[t_0,t_1]$. And similar to how the manifold $S$ may be viewed as a \emph{cobordism} from $S_0=\varphi^{-1}(t_0)$ to $S_1=\varphi^{-1}(t_1)$, the quantum state over time $(\E_1,...,\E_n)\star \rho$ is a single operator representing the dynamical evolution of $\rho_0=\rho$ into $\rho_n=(\E_n\circ \cdots \circ \E_1)(\rho)$. Moreover, similar to how the metric of $S$ picks up a Lorentzian signature as it extends over time, the state over time $(\E_1,...,\E_n)\star \rho$ may have negative eigenvalues, signifying that it is a state \emph{over} time, as opposed to a state at a single time. 

\br[\bf{Quantum states over time as a resolution of quantum quandaries?}]
If $\bold{Hilb}$, $\bold{Set}$ and $\bold{nCob}$ denote the categories consisting of linear maps between Hilbert spaces, functions between sets, and cobordisms between $n$-manifolds respectively, then the aforementioned analogy between cobordisms and quantum states over time is motivated by Ref.~\cite{Ba06}, where it is stated that certain mysterious features of quantum theory, such as local realism and the no-cloning theorem, "...only seem puzzling when we try to treat $\bold{Hilb}$ as analogous to $\bold{Set}$ rather than $\bold{nCob}$, so that quantum theory will make more sense when regarded as part of a theory of spacetime.". In the approach taken here however, it is not the category $\bold{Hilb}$ which we are treating in an analogous manner to $\bold{nCob}$, but rather, a closely related category $\bold{QInf}$, where the objects are quantum states, and the morphisms are quantum states over time. \hspace{2.71cm}$\triangle$
\er

If a system of quantum states over time satisfies the condition
\be \label{PRXPGTR767}
(\E_1,...,\E_n)\star \rho=(\E_n\circ \tr)\star \Big((\E_1,...,\E_{n-1})\star \rho\Big)
\ee
for all $n$-chains $(\E_1,...,\E_n)$ and virtual states $\rho$, then the system is said to be a \emph{propagating system of states over time}. It then follows that a propagating system of states over time is then completely determined by its assignment on pairs 
\[
(\E,\rho)\longmapsto \E\star \rho
\]
for all channels $\E$ and virtual states $\rho$. Such an assignment on pairs $(\E,\rho)$ is said to be a \emph{state over time function}~\cite{HHPBS17,FuPa22,LiNg23}. While there are many examples of state over time functions (see e.g. Ref.~\cite{FuPa22a}), the state over time function given by 
 \be \label{SBLOOMX56}
 \E\star \rho=\bloom(\E)(\rho)
 \ee
 has been shown in Refs.~\cite{LiNg23,FPBC23} to be uniquely characterized by physically motivated axioms (we recall $\bloom(\E)$ is the bloom of $\E$ as given by \eqref{BLXDXF6777}). As such, we refer to $\E\star \rho$ as given by \eqref{SBLOOMX56} as the \emph{canonical state over time} associated with the channel $\E$ and the virtual state $\rho$. A propagating system of states over time is then said to be \emph{canonical} if and only if the associated state over time function $(\E,\rho)\mapsto \E\star \rho$ outputs the canonical state over time for all channels $\E$ and virtual states $\rho$. The following theorem establishes an explicit construction of a propagating system of states over time that is canonical, which by \eqref{PRXPGTR767} is necessarily unique.
 
 \bt \label{PGPGTXS757}
 Given an $n$-chain $(\E_1,...,\E_n)\in \bold{CPTP}(\A_0,...,\A_n)$ and a virtual state $\rho\in \mathcal{V}(\A_0)$, let $(\E_1,...\E_n)\star \rho\in \A_0\otimes \cdots \otimes \A_n$ be the element given by
 \[
 (\E_1,...,\E_n)\star \rho=\bloom(\E_1,...,\E_n)(\rho),
 \]
 where $\bloom(\E_1,...,\E_n)$ is the bloom of the $n$-chain as given by \eqref{BLXDFSXS71}. Then the assignment 
\be \label{CNCNLXSTX87}
(\E_1,...,\E_n,\rho)\longmapsto (\E_1,...,\E_n)\star \rho
\ee
is a propagating system of states over time which is canonical.
\et

\bprf
Let $(\E_1,...,\E_n)\in \bold{CPTP}(\A_0,...,\A_n)$ be an $n$-chain with initial virtual state $\rho\in \mathcal{V}(\A_0)$. Then
\begin{eqnarray*}
(\E_1,...,\E_n)\star \rho&=&\bloom(\E_1,...,\E_n)(\rho) \\
&\overset{\eqref{CLXFXM87}}=&\Big(\bloom(\E_n\circ \tr)\circ \cdots \circ \bloom(\E_2\circ \tr)\circ \bloom(\E_1)\Big)(\rho) \\
&=&\bloom(\E_n\circ \tr)\Big(\Big(\bloom(\E_{n-1}\circ \tr)\circ \cdots \circ \bloom(\E_2\circ \tr)\circ \bloom(\E_1)\Big)(\rho)\Big) \\
&\overset{\eqref{CLXFXM87}}=&\bloom(\E_n\circ \tr)\Big(\bloom(\E_1,...,\E_{n-1})(\rho)\Big) \\
&=&\bloom(\E_n\circ \tr)\Big((\E_1,...,\E_{n-1})\star \rho\Big) \\
&=&(\E_n\circ \tr)\star \Big((\E_1,...,\E_{n-1})\star \rho\Big).
\end{eqnarray*}
thus \eqref{PRXPGTR767} holds. It then follows that the assignment \eqref{CNCNLXSTX87} is a propagating system of states over time, and since $\E\star \rho=\bloom(\E)(\rho)$, it follows that the the assignment \eqref{CNCNLXSTX87} is also canonical.
\eprf

\section{Covariance of canonical states over time} \label{SX4}

In general relativity, one of the fundamental principles is that of general covariance, meaning that the form of equations governing physical laws should be unaltered by arbitrary smooth coordinate transformations of spacetime. More precisely, if $(x,y,z,t)$ are local coordinates on spacetime, and a physical law takes the form 
\[
\mathcal{G}(x,y,z,t)=\mathcal{T}(x,y,z,t),
\]
then the equation
\[
\mathcal{G}(x',y',z',t')=\mathcal{T}(x',y',z',t')
\]
should also hold, where $(x',y',z',t')$ is another set of coordinates on spacetime, differing from $(x,y,z,t)$ by a smooth coordinate transformation.

In the quantum setting, suppose that a system in an initial state $\rho(0)\in \mathcal{S}(\A)$ with Hamiltonian $H$ evolves according to the Liouville-von Neumann equation, so that
\be \label{LVXVNX747}
i\frac{\partial}{\partial t}\rho(t)=\Big[H,\rho(t)\Big].
\ee
If $\A$ is a matrix algebra and $U\in \A$ is unitary, then setting $\rho'=U\rho U^{\dag}$ and $H'=UHU^{\dag}$ yields
\begin{eqnarray*}
i\frac{\partial}{\partial t}\rho'(t)&=&i\frac{\partial}{\partial t}\Big(U\rho(t)U^{\dag}\Big) \\
&=&U\Big(i\frac{\partial}{\partial t}\rho(t)\Big)U^{\dag} \\
&\overset{\eqref{LVXVNX747}}=&U(H\rho(t)-\rho(t)H)U^{\dag} \\
&=&(UHU^{\dag})(U\rho(t)U^{\dag})-(U\rho(t)U^{\dag})(UHU^{\dag}) \\
&=&\Big[H',\rho'(t)\Big],
\end{eqnarray*}
so that
\be  \label{LVXVNX7479}
i\frac{\partial}{\partial t}\rho'(t)=\Big[H',\rho'(t)\Big].
\ee
As such, if we view $\Ad_{U}:\A\to \A$ as a coordinate transformation corresponding to a different choice of an orthonormal basis in which to from a matrix representation of our system, then it follows that the Liouville-von Neumann equation is \emph{unitarily} covariant, i.e., applying a unitary transformation to both sides of equation \eqref{LVXVNX747} yields equation \eqref{LVXVNX7479}\footnote{Certainly any first-order linear differential equation in $\rho$ is also unitarily covariant.}. 

In terms of quantum states over time, we can view the Liouville-von Neumann equation \eqref{LVXVNX747} as determining a canonical state over time
\[
(\E_1,...,\E_n)\star \rho,
\]
where $\rho=\rho(0)$ and $\E_k:\A\to \A$ is the unitary transformation given by 
\[
\E_k(A)=e^{-iHt_k}Ae^{iHt_k}
\]
for all $A\in \A$ and for all $k\in [n]$. In such a case, we view the continuous time parameter $t$ as a smooth approximation of the discrete time-steps $\Delta t=t_{m+1}-t_{m}$ associated with the $n$-chain $(\E_1,...,\E_n)$ for large $n$. Moreover, if $U\in \A$ is a unitary operator corresponding to a coordinate transformation $\text{Ad}_U:\A\to \A$, then for all $k\in [n]$ the operator $\E_k$ transforms to the operator $\E_k'=\text{Ad}_U\circ \E_{k}\circ \text{Ad}_{U^{\dag}}$, thus for all $A'\in \A$ we have
\begin{eqnarray*}
\E_k'(A')&=&\Big(\text{Ad}_U\circ \E_{k}\circ \text{Ad}_{U^{\dag}}\Big)(A') \\
&=&\text{Ad}_U\Big(\E_k(U^{\dag}A'U)\Big) \\
&=&\text{Ad}_U\Big(e^{-iHt_k}U^{\dag}A'Ue^{iHt_k}\Big) \\
&=&\Big(Ue^{-iHt_k}U^{\dag}\Big)A'\Big(Ue^{iHt_k}U^{\dag}\Big) \\
&=&e^{-i(UHU^{\dag})t_k}A'e^{i(UHU^{\dag})t_k} \\
&=&e^{-iH't_k}A'e^{iH't_k}, 
\end{eqnarray*}
where again $H'=UHU^{\dag}$ is the associated transformation of the Hamiltonian $H$. It then follows that the transformed Liouville-von Neumann equation \eqref{LVXVNX7479} determines the canonical state over time $(\E_1',...,\E_n')\star \rho'$, where $\rho'=U\rho U^{\dag}$. If the theory of quantum states over time is to be covariant, then it should follow that the canonical state over time $(\E_1,...,\E_1)\star \rho$ associated with the Liouville-von Neumann equation \eqref{LVXVNX747} transforms under the unitary transformation $\text{Ad}_U\otimes \cdots \otimes \text{Ad}_U$ to the canonical state over time $(\E_1',...,\E_n')\star \rho'$ associated with the transformed Liouville-von Neumann equation \eqref{LVXVNX7479}, i.e., 
\be \label{CVXSTXSX967}
(\text{Ad}_U\otimes \cdots \otimes \text{Ad}_U)\Big((\E_1,...,\E_n)\star \rho\Big)=(\E_1',...,\E_n')\star \rho'.
\ee

In this section, we establish Theorem~\ref{MTX87}, which states that not only does equation \eqref{CVXSTXSX967} indeed hold, but that equation \eqref{CVXSTXSX967} is a special case of a more general covariance relation between canonical states over time. For this, suppose that a quantum system in an initial state $\rho\in \mathcal{S}(\A_0)$ is to evolve according to an $n$-chain $(\E_1,...,\E_n)\in \bold{CPTP}(\A_0,...,\A_n)$. Then for every $(n+1)$-tuple
\[
(\phi_0:\A_0\lra \A_0'\,,\ldots,\,\phi_n:\A_n\lra \A_n')
\]
of $*$-isomorphisms, one may equivalently describe the dynamical scenario as the initial state $\rho'=\phi_0(\rho)$ evolving according to the $n$-chain $(\E_1',...,\E_n')\in \bold{CPTP}(\A_0',...,\A_n')$, where now $\E_k'=\phi_k\circ \E_k\circ \phi_{k-1}^{-1}$ for all $k\in [n]$. As these two mathematical descriptions of such a dynamical scenario are equivalent, we should expect that the canonical state over time $(\E_1,...,\E_n)\star \rho$ transforms to the canonical state over time $(\E_1',...,\E_n')\star \rho'$ under the $*$-isomorphism $(\phi_0\otimes \cdots \otimes \phi_n)$. The following theorem states that this is indeed the case, which we view as a general covariance property for canonical states over time.

\bt[Covariance of canonical states over time] \label{MTX87}
Let $\star$ be the operation corresponding to the canonical propagating system of states over time, let $(\E_1,...,\E_n)\in \bold{CPTP}(\A_0,...,\A_n)$ be an $n$-chain, and let $\rho\in \mathcal{V}(\A_0)$. Then for every $(n+1)$-tuple $(\phi_0:\A_0\to \A_0',...,\phi_n:\A_n\to \A_n')$ of $*$-isomorphisms,
\be \label{CVXSXT367}
(\phi_0\otimes \cdots \otimes \phi_n)\Big((\E_1,...,\E_n)\star \rho\Big)=(\E_1',...,\E_n')\star \rho',
\ee
where $\E_k'=\phi_k\circ \E_k\circ \phi_{k-1}^{-1}$ for all $k\in [n]$ and $\rho'=\phi_0(\rho)$.
\et

The theorem is essentially a direct corollary of Theorem~\ref{CVXNBLX767}, which appears in the appendix.

\bprf[Proof of Theorem~\ref{MTX87}]
Since $\E_k'=\phi_i\circ \E_k\circ \phi_{k-1}^{-1}$ for all $k\in [n]$, it follows from Theorem~\ref{CVXNBLX767} that
\be \label{EXQ757}
(\phi_0\otimes \cdots \otimes \phi_n)\circ \bloom(\E_1,...,\E_n)=\bloom(\E_1',...,\E_n')\circ \phi_0.
\ee
We then have
\begin{eqnarray*}
(\phi_0\otimes \cdots \otimes \phi_n)\Big((\E_1,...,\E_n)\star \rho\Big)&=&(\phi_0\otimes \cdots \otimes \phi_n)\Big(\bloom(\E_1,...,\E_n)(\rho)\Big) \\
&=&\Big((\phi_0\otimes \cdots \otimes \phi_n)\circ \bloom(\E_1,...,\E_n)\Big)(\rho) \\
&\overset{\eqref{EXQ757}}=&\Big(\bloom(\E_1',...,\E_n')\circ \phi_0\Big)(\rho) \\
&=&\bloom(\E_1',...,\E_n')\Big(\phi_0(\rho)\Big) \\
&=&\bloom(\E_1',...,\E_n')(\rho') \\
&=&(\E_1',...,\E_n')\star \rho',
\end{eqnarray*}
as desired.
\eprf

In the next section, we address the covariance of the \emph{dynamical quantum Bayes' rule} introduced in Ref.~\cite{FuPa22a}.

\section{Covariance of the dynamical quantum Bayes' rule} \label{SX5}

Let $X$ and $Y$ be discrete random variables with probability mass functions $\mathbb{P}(x)$ and $\mathbb{P}(y)$. The classical \emph{Bayes' rule} may be formulated as the equality
\be \label{BRX81}
\mathbb{P}(x)\mathbb{P}(y|x)=\mathbb{P}(y)\mathbb{P}(x|y)
\ee
for all $x\in X$ and $y\in Y$. A dynamical generalization of Bayes' rule using quantum states over time first appeared  in \cite{FuPa22a}, where it was shown that many fundamental notions associated with time-reversal in quantum theory --- such as the state-update rule, the 2-state vector formalism of Reznik and Aharonov \cite{ReAh95}, the Petz recovery map \cite{OhPe93}, the generalized conditional expectations of Tsang \cite{Ts22}, and the quantum Bayes' theorem of Leifer and Spekkens \cite{LeSp13} --- may all be formulated in terms of such a dynamical quantum Bayes' rule\footnote{In Ref. \cite{FuPa22}, the quantum Bayes' rule is defined with respect to general states over time, while we only consider canonical states over time as given by \eqref{SBLOOMX56}.}. 

To motivate the dynamical quantum Bayes' rule, let $\vartheta:X\times Y\to [0,1]$ be the joint distribution corresponding to first observing the output of $X$ and then observing the output of $Y$. It then follows that
\[
\vartheta(x,y)=\mathbb{P}(x)\mathbb{P}(y|x),
\]
as obtaining the outcome $(x,y)$ corresponds to first obtaining the outcome $x$, and then obtaining the outcome $y$ \emph{given} $x$. Similarly, there is a joint distribution $\vartheta^*:Y\times X\to [0,1]$ associated with reversing the order of observation (i.e., by observing $Y$ before $X$), so that
\[
\vartheta^*(y,x)=\mathbb{P}(y)\mathbb{P}(x|y).
\]
Bayes' rule \eqref{BRX81} may then be reformulated as
\be \label{NBRX71}
\vartheta=\gamma \left(\vartheta^*\right),
\ee
where $\gamma:\Pi(Y\times X)\lra \Pi(X\times Y)$ is the natural isomorphism between probability distributions on $Y\times X$ and probability distributions on $X\times Y$ given by
\[
\gamma(q)(x,y)=q(y,x)
\]
for all $q\in \Pi(Y\times X)$. Moreover, as $\vartheta^*$ is viewed as the distribution associated with the time-reversal of the the process associated with the distribution $\vartheta$, we view \eqref{NBRX71} as a dynamical formulation of Bayes' rule, so that $\gamma$ is then viewed as a time-reversal map.

To generalize Bayes' rule to the quantum domain, we recast the dynamical Bayes' rule \eqref{NBRX71} in terms of CPTP dynamics. For this, we associate the conditional probabilities $\mathbb{P}(y|x)$ and $\mathbb{P}(x|y)$ with classical channels $\E:\mathbb{C}^X\to \mathbb{C}^Y$ and $\ov \E:\mathbb{C}^Y\to \mathbb{C}^X$ respectively (see Remark~\ref{RM7X1787}), and we associate the distributions $\mathbb{P}(x)$ and $\mathbb{P}(y)$ with states $\rho\in \mathcal{S}(\mathbb{C}^X)$ and $\sigma\in \mathcal{S}(\mathbb{C}^Y)$, so that $\sigma=\E(\rho)$. It then follows that $\vartheta\in \Pi(X\times Y)$ corresponds to the state over time $\E\star \rho$, while $\vartheta^*\in \Pi(Y\times X)$ corresponds to the state over time $\ov \E\star \sigma$. The dynamical Bayes' rule \eqref{NBRX71} is then equivalent to the equation
\be \label{QBR737X}
\E\star \rho = \gamma \left(\ov \E \star \sigma\right),
\ee
where now $\gamma:\mathbb{C}^Y\otimes \mathbb{C}^X\to \mathbb{C}^X\otimes \mathbb{C}^Y$ is the lexicographic swap isomorphism. Note however that equation \eqref{QBR737X} makes sense for any channels $\E\in \bold{CPTP}(\A,\B)$ and $\ov \E\in \bold{CPTP}(\B,\A)$ with $\A$ and $\B$ arbitrary finite-dimensional $C^*$-algebras. This motivates the following:

\bd
Let $\A$ and $\B$ be finite-dimensional $C^*$-algebras, let $\E\in \bold{CPTP}(\A,\B)$, and let $\rho\in \mathcal{S}(\A)$ be a state. A map $\ov \E\in \bold{CPTP}(\B,\A)$ is then said to satisfy the \define{quantum Bayes' rule} with respect to the pair $(\E,\rho)$ if and only if
\be \label{CVXBRXS187}
\E\star \rho = \gamma \left(\ov \E \star \E(\rho)\right),
\ee
where $\gamma:\B\otimes \A\to \A\otimes \B$ is the lexicographic swap isomorphism.
\ed 

Given a pair $(\phi:\A\to \A', \psi:\A\to \B')$ of $*$-isomorphisms, the maps $\E\in \bold{CPTP}(\A,\B)$ and $\ov \E\in \bold{CPTP}(\B,\A)$ in the quantum Bayes' rule \eqref{CVXBRXS187} transform to the maps $\E'=\psi\circ \E\circ \phi^{-1} \in \bold{CPTP}(\A',\B')$ and $\ov \E'=\phi\circ \ov \E\circ \psi^{-1}\in \bold{CPTP}(\B',\A')$, while the state $\rho\in \mathcal{S}(\A)$ transforms to $\rho'=\phi(\rho)\in \mathcal{S}(\A')$. The following theorem then states that the dynamical quantum Bayes' rule is covariant, i.e., \eqref{CVXBRXS187} still holds after replacing $\E$, $\ov \E$, and $\rho$ by their primed counterparts, namely, $\E'$, $\ov \E'$ and $\rho'$.   

\bt[Covariance of the Quantum Bayes' Rule] \label{CVQBXR57}
Let $\E\in \bold{CPTP}(\A,\B)$, $\rho\in \mathcal{S}(\A)$, suppose $\overline{\E}\in \bold{CPTP}(\B,\A)$ satisfies the quantum Bayes' rule \eqref{CVXBRXS187} with respect to $(\E,\rho)$, and let $\phi:\A\to \A'$ and $\psi:\B\to \B'$ be $*$-isomorphisms. Then
\be \label{CVXBRXS181}
\E'\star \rho'=\gamma'\left(\overline{\E}'\star \E'(\rho')\right),
\ee
where $\E'=\psi\circ \E\circ \phi^{-1}$, $\overline{\E}'=\phi\circ \overline{\E}\circ \psi^{-1}$, $\rho'=\phi(\rho)$, and $\gamma':\B'\otimes \A'\to \A'\otimes \B'$ is the lexicographic swap isomorphism.
\et

Before giving a proof we first prove the following:

\blem
Let $\phi:\A\to \A'$ and $\psi:\B\to \B'$ be $*$-isomorphisms, and let $\gamma:\B\otimes \A\to \A\otimes \B$ and $\gamma':\B'\otimes \A'\to \A'\otimes \B'$ be the lexicographic swap isomorphisms. Then
\be \label{EQX371}
(\phi\otimes \psi)\circ \gamma=\gamma'\circ (\psi\otimes \phi).
\ee
\elem

\bprf
Let $A\in \A$ and $B\in \B$. Then
\begin{eqnarray*}
\Big((\phi\otimes \psi)\circ \gamma\Big)(B\otimes A)&=&(\phi\otimes \psi)(A\otimes B) \\
&=&\phi(A)\otimes \psi(B) \\
&=&\gamma'(\psi(B)\otimes \phi(A)) \\
&=&\Big(\gamma'\circ (\psi\otimes \phi)\Big)(B\otimes A).
\end{eqnarray*}
The statement then follows from the linearity of the maps $(\phi\otimes \psi)\circ \gamma$ and $\gamma'\circ (\psi\otimes \phi)$.
\eprf

\bprf[Proof of Theorem~\ref{CVQBXR57}]
Let $\phi:\A\to \A'$ and $\psi:\B\to \B'$ be a $*$-isomorphisms, $\E'=\psi\circ \E\circ \phi^{-1}$, $\overline{\E}'=\phi\circ \overline{\E}\circ \psi^{-1}$, $\rho'=\phi(\rho)$, and let $\gamma':\B'\otimes \A'\to \A'\otimes \B'$ be the lexicographic swap isomorphism. Then covariance of states over time \eqref{CVXSXT367} yields
\be  \label{EQX373}
(\phi\otimes \psi)(\E\star \rho)=\E'\star \phi(\rho) \qquad \& \qquad (\psi\otimes \phi)\left(\overline{\E}\star \E(\rho)\right)=\overline{\E}'\star \psi(\E(\rho)).
\ee
We then have
\begin{eqnarray*}
\E'\star \phi(\rho)&\overset{\eqref{EQX373}}=&(\phi\otimes \psi)(\E\star \rho) \\
&\overset{\eqref{CVXBRXS187}}=&(\phi\otimes \psi)\left(\gamma\left(\overline{\E}\star \E(\rho)\right)\right) \\
&=&\left(\phi\otimes \psi)\circ \gamma\right)\left(\overline{\E}\star \E(\rho)\right) \\
&\overset{\eqref{EQX371}}=&\left(\gamma'\circ (\psi\otimes \phi)\right)\left(\overline{\E}\star \E(\rho)\right) \\
&=&\gamma'\left((\psi\otimes \phi)\left(\overline{\E}\star \E(\rho)\right)\right) \\
&\overset{\eqref{EQX373}}=&\gamma'\left(\overline{\E}'\star \psi(\E(\rho))\right) \\
&=&\gamma'\left(\overline{\E}'\star \E'(\phi(\rho))\right) \\
&=&\gamma'\left(\overline{\E}'\star \E'(\rho')\right),
\end{eqnarray*}
thus \eqref{CVXBRXS181} holds.
\eprf

\section{Concluding remarks} \label{SX6}

In this work, we formulated a precise mathematical notion of general covariance for the theory of quantum states over time. We also defined the propagator condition \eqref{PRXPGTR767} for systems of quantum states over time, which combined with the recent uniqueness results for state over time functions appearing in \cite{LiNg23,FPBC23}, yields a canonical system of states over time associated with quantum systems evolving under general CPTP dynamics. We then proved that such a canonical system of states over time is indeed covariant, and we were also able to prove that the dynamical quantum Bayes' rule is generally covariant as well.

While the theory of quantum states over time has yet to develop fundamental dynamical laws analogous to general relativity, it is our hope that the establishment of general covariance will help provide a foundation for the search of physical laws formulated in terms of quantum states over time. In particular, such a physical law should take the form of  
\be \label{DYNXLX45}
\mathfrak{G}(\rho,\E_1,...,\E_n)=\mathfrak{T}(\rho,\E_1,...,\E_n),
\ee
where $(\E_1,...,\E_n)$ is an $n$-chain to be determined via \eqref{DYNXLX45} and initial state $\rho$. If equation \eqref{DYNXLX45} is taking place in $\A_0\otimes \cdots \otimes \A_n$, then similar to the dynamical quantum Bayes' rule \eqref{CVXBRXS187}, such a law should transform covariantly with the associated quantum states over time. More precisely, for every $(n+1)$-tuple of $*$-isomorphisms $(\phi_0:\A_0\to \A_0',...,\phi_n:\A_n\to \A_n')$, we should have
\[
(\phi_0\otimes \cdots \otimes \phi_n)\Big(\mathfrak{G}(\rho,\E_1,...,\E_n)\Big)=\mathfrak{G}(\rho',\E_1',...,\E_n')
\]
and
\[
(\phi_0\otimes \cdots \otimes \phi_n)\Big(\mathfrak{T}(\rho,\E_1,...,\E_n)\Big)=\mathfrak{T}(\rho',\E_1',...,\E_n'),
\]
where $\rho'=\phi_0(\rho)$ and $\E_{k}'=\phi_{k}\circ \E_k\circ \phi_{k-1}^{-1}$ for all $k\in [n]$. In such a case, equation \eqref{DYNXLX45} holds if and only if the equation
\[
\mathfrak{G}(\rho',\E_1',...,\E_n')=\mathfrak{T}(\rho',\E_1',...,\E_n')
\]
holds as well, which we view as general covariance for the (hypothetical) physical law associated with \eqref{DYNXLX45}. 

\appendix

\section{Covariance of canonical broadcasting}

\bn[Covariance of canonical broadcasting]
Let $\phi:\A\to \A'$ be a $*$-isomorphism. Then
\be \label{BDXCVX783}
\mathfrak{B}_{\A'}\circ \phi=(\phi\otimes \phi)\circ \mathfrak{B}_{\A} \hspace{1mm}.
\ee
\en

\bprf
First, note that \eqref{BDXCVX783} is equivalent to the equation
\be \label{BDCASXT97}
\phi^{-1}\circ \mathfrak{B}_{\A'}^*\circ (\phi\otimes \phi)=\mathfrak{B}_{\A}^*,
\ee
where $\mathfrak{B}_{\A}^*$ and $\mathfrak{B}_{\A'}^*$ denote the Hilbert-Schmidt adjoints of $\mathfrak{B}_{\A}$ and $\mathfrak{B}_{\A'}$. As such, the statement is proved if we show \eqref{BDCASXT97} holds. Indeed, for all $A_1,A_2\in \A$ we have
\begin{eqnarray*}
\left(\phi^{-1}\circ \mathfrak{B}_{\A'}^*\circ (\phi\otimes \phi)\right)(A_1\otimes A_2)&=&\phi^{-1}\left(\mathfrak{B}_{\A'}^*(\phi(A_1)\otimes \phi(A_2))\right) \\
&=&\phi^{-1}\left(\frac{1}{2}(\mu_{\A'}(\phi(A_1)\otimes \phi(A_2))+\widetilde{\mu}_{\A'}(\phi(A_1)\otimes \phi(A_2)))\right) \\
&=&\phi^{-1}\left(\frac{1}{2}(\phi(A_1)\phi(A_2)+\phi(A_2)\phi(A_1))\right) \\
&=&\frac{1}{2}(A_1A_2+A_2A_1) \\
&=&\frac{1}{2}(\mu_{\A}(A_1\otimes A_2)+\widetilde{\mu}_{\A}(A_1\otimes A_2)) \\
&=&\mathfrak{B}_{\A}^*(A_1\otimes A_2),
\end{eqnarray*}
thus \eqref{BDCASXT97} holds.
\eprf

\section{Covariance of bloom maps}

\bt[Covariance of bloom maps] \label{CVXNCE71}
Let $\phi:\A\to \A'$ and $\psi:\B\to \B'$ be $*$-isomorphisms, and suppose $\E\in \bold{TP}(\A,\B)$ and $\E'\in \bold{TP}(\A',\B')$ are such that $\psi\circ \E=\E'\circ \phi$. Then
\be \label{CVX1BM797}
(\phi\otimes \psi)\circ \bloom(\E)=\bloom(\E')\circ \phi \hspace{1mm}.
\ee
\et

\bprf
Indeed,
\begin{eqnarray*}
(\phi\otimes \psi)\circ \bloom(\E)&=&(\phi\otimes \psi)\circ (\id_{\A}\otimes \E)\circ \mathfrak{B}_{\A} \\
&=&\left(\phi\otimes (\psi\circ \E)\right) \circ \mathfrak{B}_{\A} \\
&=&\left(\phi\otimes (\E'\circ \phi)\right) \circ \mathfrak{B}_{\A} \\
&=&\left((\id_{\A'}\circ \phi)\otimes (\E'\circ \phi)\right) \circ \mathfrak{B}_{\A} \\
&=&(\id_{\A'}\otimes \E')\circ (\phi\otimes \phi)\circ \mathfrak{B}_{\A} \\
&\overset{\eqref{BDXCVX783}}=&(\id_{\A'}\otimes \E')\circ \mathfrak{B}_{\A}\circ \phi \\
&=&\bloom(\E')\circ \phi,
\end{eqnarray*}
as desired.
\eprf

\section{Covariance of 2-blooms}

\bd
Let $(\phi:\A\to \A', \hspace{0.5mm}\psi:\B\to \B',\hspace{0.5mm} \vartheta:\C\to \C')$ be a 3-tuple of $*$-isomorphisms. Two 2-chains $(\E,\F)\in \bold{TP}(\A,\B,\C)$ and $(\E',\F')\in \bold{TP}(\A',\B',\C')$ are said to be \define{$(\phi,\psi,\vartheta)$-equivalent} if and only if $\psi\circ \E=\E'\circ \phi$ and $\vartheta\circ \F=\F'\circ \psi$.
\ed

\blem \label{LXMAX67}
Suppose $(\E,\F)\in \bold{TP}(\A,\B,\C)$ and $(\E',\F')\in \bold{TP}(\A',\B',\C')$ are $(\phi,\psi,\vartheta)$-equivalent. Then
\be \label{LMX2BM19}
(\psi\otimes \vartheta)\circ \bloom(\F)\circ \E=\bloom(\F')\circ \E'\circ \phi \hspace{1mm}.
\ee
\elem

\bprf
Since $(\E,\F)\in \bold{TP}(\A,\B,\C)$ and $(\E',\F')\in \bold{TP}(\A',\B',\C')$ are $(\phi,\psi,\vartheta)$-equivalent we have
\be \label{2EQX1}
\psi\circ \E=\E'\circ \phi
\ee
and
\be \label{2EQX2}
\vartheta\circ \F=\F'\circ \psi,
\ee
thus
\begin{eqnarray*}
(\psi\otimes \vartheta)\circ \bloom(\F)\circ \E&=&(\psi\otimes \vartheta)\circ (\id_{\B}\otimes \F)\circ \mathfrak{B}_{\B}\circ \E \\
&=&(\psi\otimes (\vartheta\circ \F))\circ \mathfrak{B}_{\B}\circ \E \\
&\overset{\eqref{2EQX2}}=&(\psi\otimes (\F'\circ \psi))\circ \mathfrak{B}_{\B}\circ \E \\
&=&(\id_{\B'}\otimes \F')\circ (\psi\otimes \psi)\circ \mathfrak{B}_{\B}\circ \E \\
&\overset{\eqref{BDXCVX783}}=&(\id_{\B'}\otimes \F')\circ \mathfrak{B}_{\B'}\circ \psi\circ \E \\
&\overset{\eqref{2EQX1}}=&\bloom(\F')\circ \E'\circ \phi,
\end{eqnarray*}
as desired.
\eprf

\bd
The \define{bloom} of a 2-chain $(\E,\F)\in \bold{TP}(\A,\B,\C)$ is the map $\bloom(\E,\F)\in \bold{TP}(\A,\A\otimes \B\otimes \C)$ given by
\[
\bloom(\E,\F)=\bloom(\bloom(\F)\circ \E).
\]
\ed

\bt[Covariance of 2-blooms]
Suppose $(\E,\F)\in \bold{TP}(\A,\B,\C)$ and $(\E',\F')\in \bold{TP}(\A',\B',\C')$ are $(\phi,\psi,\vartheta)$-equivalent. Then
\be
(\phi\otimes \psi\otimes \vartheta)\circ \bloom(\E,\F)=\bloom(\E',\F')\circ \phi  \hspace{1mm}.
\ee
\et

\bprf
Let $\widetilde{\E}=\bloom(\F)\circ \E$, $\widetilde{\E}'=\bloom(\F')\circ \E'$, $\widetilde{\phi}=\phi$ and $\widetilde{\psi}=\psi\otimes \vartheta$. By \eqref{LMX2BM19} we have $\widetilde{\psi}\circ \widetilde{\E}=\widetilde{\E}'\circ \widetilde{\phi}$, thus by \eqref{CVX1BM797} we have $(\widetilde{\phi}\otimes \widetilde{\psi})\circ \bloom(\widetilde{\E})=\bloom(\widetilde{\E}')\circ \widetilde{\phi}$, which when written without the substitutions yields
\be \label{EQX7971}
(\phi\otimes \psi\otimes \vartheta)\circ \bloom(\bloom(\F)\circ \E)=\bloom(\bloom(\F')\circ \E')\circ \phi \hspace{1mm}.
\ee 
We then have
\begin{eqnarray*}
(\phi\otimes \psi\otimes \vartheta)\circ \bloom(\E,\F)&=&(\phi\otimes \psi\otimes \vartheta)\circ \bloom(\bloom(\F)\circ \E) \\
&\overset{\eqref{EQX7971}}=&\bloom(\bloom(\F')\circ \E')\circ \phi \\
&=&\bloom(\E',\F')\circ \phi,
\end{eqnarray*}
as desired.
\eprf

\section{Covariance of $n$-blooms}

\bd \label{PHIXNEQX87}
Let $\left(\phi_0:\A_0\to \A_0',...,\phi_n:\A_n\to \A_n'\right)$ be an $(n+1)$-tuple of $*$-isomorphisms. Two $n$-chains $(\E_1,...,\E_n)\in \bold{TP}(\A_0,...,\A_n)$ and $(\E_1',...,\E_n')\in \bold{TP}(\A_0',...,\A_n')$ are said to be \define{$(\phi_0,...,\phi_n)$-equivalent} if and only if for all $i\in [n]$ we have $\phi_i\circ \E_i=\E_i'\circ \phi_{i-1}$.
\ed

\blem \label{2XQEV381}
Suppose $(\E_1,...,\E_n)\in \bold{TP}(\A_0,...,\A_n)$ and $(\E_1',...,\E_n')\in \bold{TP}(\A_0',...,\A_n')$ are $(\phi_0,...,\phi_n)$-equivalent. Then $(\E_1,\bloom(\E_3,...,\E_n)\circ \E_2)$ and $(\E_1',\bloom(\E_3',...,\E_n')\circ \E_2')$ are $(\phi_0,\phi_1,\phi_2\otimes \cdots \otimes \phi_n)$-equivalent.
\elem

\bprf
We use induction on $n\geq 2$. The case $n=2$ follows by definition. Now suppose the result holds for $n=m-1$ with $m>2$. Since we already have $\phi_1\circ \E_1=\E_1'\circ \phi_0$, the result then follows once we show
\be \label{EXQ737}
(\phi_2\otimes \cdots \otimes \phi_m)\circ \bloom(\E_3,...,\E_m)\circ \E_2=\bloom(\E_3',...,\E_m')\circ \E_2' \circ \phi_1.
\ee
So let $\widetilde{\E}_3=\bloom(\E_4,...,\E_m)\circ \E_3$, $\widetilde{\E}_3'=\bloom(\E_4',...,\E_m')\circ \E_3'$, and let $\widetilde{\phi}_3=\phi_3\otimes \cdots \otimes \phi_m$ (if $m=3$ then $\widetilde{\E}_3=\E_3$ and $\widetilde{\phi}_3=\phi_3$). Our inductive assumption applied to the $(m-1)$-chain $(\E_2,...,\E_m)$ yields
\be \label{IDXSMX67}
\widetilde{\phi}_3\circ \widetilde{\E}_3=\widetilde{\E}_3'\circ \phi_2,
\ee
thus by Theorem~\ref{CVXNCE71} we have $(\phi_2\otimes \widetilde{\phi_3})\circ \bloom(\widetilde{\E}_3)=\bloom(\widetilde{\E}_3')\circ \phi_2$, i.e.,
\be \label{2TXCVX771}
(\phi_2\otimes \cdots \otimes \phi_m)\circ \bloom(\bloom(\E_4,...,\E_m)\circ \E_3)=\bloom(\bloom(\E_4',...,\E_m')\circ \E_3')\circ \phi_2.
\ee
We then have
\begin{eqnarray*}
(\phi_2\otimes \cdots \otimes \phi_m)\circ \bloom(\E_3,...,\E_m)\circ \E_2&=&(\phi_2\otimes \cdots \otimes \phi_m)\circ \bloom(\bloom(\E_4,...,\E_m)\circ \E_3)\circ \E_2 \\
&\overset{\eqref{2TXCVX771}}=&\bloom(\bloom(\E_4',...,\E_m')\circ \E_3')\circ \phi_2\circ \E_2 \\
&=&\bloom(\E_3',\E_4',...,\E_m')\circ \E_2'\circ \phi_1,
\end{eqnarray*}
where the final equality follows from the fact that $\phi_2\circ \E_2=\E_2'\circ \phi_1$, thus concluding the proof.
\eprf

\blem \label{LMXNBM19}
Suppose $(\E_1,...,\E_n)\in \bold{TP}(\A_0,...,\A_n)$ and $(\E_1',...,\E_n')\in \bold{TP}(\A_0',...,\A_n')$ are $(\phi_0,...,\phi_n)$-equivalent. Then 
\be \label{LMXNBM197}
(\phi_1\otimes \cdots \otimes \phi_n)\circ \bloom(\E_2,...,\E_n)\circ \E_1=\bloom(\E_2',...,\E_n')\circ \E_1'\circ \phi_0 \hspace{1mm}.
\ee
\elem

\bprf
Let $\E=\E_1$ and $\F=\bloom(\E_3,...,\E_n)\circ \E_2$, so that $\E'=\E_1'$ and $\F'=\bloom(\E_3',...,\E_n')\circ \E_2'$. By Lemma~\ref{2XQEV381} it follows that $(\E,\F)$ and $(\E',\F')$ are $(\phi_0,\phi_1,\phi_2\otimes \cdots \otimes \phi_n)$-equivalent, thus by Lemma~\ref{LXMAX67} we have
\[
(\phi_1\otimes \phi_2\otimes \cdots \otimes \phi_n)\circ \bloom(\F)\circ \E=\bloom(\F')\circ \E'\circ \phi_0 
\]
which when written without the substitutions is precisely \eqref{LMXNBM197}, as desired.
\eprf

\bt[Covariance of $n$-blooms] \label{CVXNBLX767}
Suppose $(\E_1,...,\E_n)\in \bold{TP}(\A_0,...,\A_n)$ and $(\E_1',...,\E_n')\in \bold{TP}(\A_0',...,\A_n')$ are $(\phi_0,...,\phi_n)$-equivalent. Then
\be \label{CVXNCE77}
(\phi_0\otimes \cdots \otimes \phi_n)\circ \bloom(\E_1,...,\E_n)=\bloom(\E_1',...,\E_n')\circ \phi_0 \hspace{1mm}.
\ee
\et

\bprf
Let $\E=\bloom(\E_2,...,\E_n)\circ \E_1$, $\E'=\bloom(\E_2',...,\E_n')\circ \E_1'$, $\phi=\phi_0$, and $\psi=\phi_1\otimes \cdots \otimes \phi_n$. By Lemma~\ref{LMXNBM19} equation \eqref{LMXNBM197} holds, i.e., $\psi\circ \E=\E'\circ \phi$. It then follows from Theorem~\ref{CVXNCE71} that $(\phi\otimes \psi)\circ \bloom(\E)=\bloom(\E')\circ \phi$, which when written without the substitutions yields
\[
(\phi_0\otimes \cdots \otimes \phi_n)\circ \bloom(\bloom(\E_2,...,\E_n)\circ \E_1)=\bloom(\bloom(\E_2',...,\E_n')\circ \E_1')\circ \phi_0.
\]
And since $\bloom(\E_1,...,\E_n)=\bloom(\bloom(\E_2,...,\E_n)\circ \E_1)$ and $\bloom(\E_1',...,\E_n')=\bloom(\bloom(\E_2',...,\E_n')\circ \E_1')$, the result follows.
\eprf

\addcontentsline{toc}{section}{\numberline{}Bibliography}
\bibliographystyle{plain}
\bibliography{CVX}

\end{document}